\documentclass[12pt,a4paper]{article}
\usepackage[utf8]{inputenc}
\usepackage{amsmath}
\usepackage{placeins}
\usepackage{url}
\usepackage{natbib}
\usepackage{color,xcolor,setspace,geometry}
\setcitestyle{authoryear,open={(},close={)}}
\usepackage{cite}
\usepackage{soul}
\usepackage{bbm}
\usepackage{dsfont}
\usepackage{tablefootnote}
\usepackage{soul}
\usepackage{xcolor}
\usepackage{cite}
\sethlcolor{yellow} 
\soulregister\citep7
\soulregister\cite7
\soulregister\citet7
\soulregister\citealp7
\soulregister\ref7
\usepackage{amsfonts}
\usepackage{tikz}
\usetikzlibrary{shapes}
\usetikzlibrary{plotmarks}
\usetikzlibrary{tikzmark,matrix,calc}
\usetikzlibrary{arrows.meta,calc,decorations.markings,math,arrows.meta}
\usepackage{amssymb}
\usepackage{multirow}
\usepackage{todonotes}
\usepackage{threeparttable}
\usepackage{graphicx}
\usepackage{array}
\usepackage{arydshln}
\usepackage[final]{changes}
\usepackage{lineno}
\providecommand{\keywords}[1]
{
  \small	
  \textbf{\textit{Keywords---}} #1
}

\usepackage{natbib}
\usepackage{datatool}
\usepackage{etoolbox}

\DTLnewdb{bibnotes}

\def\bibnote#1#2{
  \DTLnewrow{bibnotes}
  \DTLnewdbentry{bibnotes}{mylabel}{#1}
  \DTLnewdbentry{bibnotes}{mynote}{#2}
}

\makeatletter
\patchcmd{\@lbibitem}
  {\item[\hfil\NAT@anchor{#2}{\NAT@num}]}
  {
    \item[\hfil\NAT@anchor{#2}{\NAT@num}]
    \DTLforeach[\DTLiseq{\mylabel}{#2}]{bibnotes}{\mylabel=mylabel,\mynote=mynote}{\text{\mynote}}
  }{}{\message{^^JPatching failed^^J}}

\makeatother
\begin{document}
\begin{spacing}{1.35}
\listofchanges
\newpage
  
\title{Bettors' reaction to match dynamics --- Evidence from in-game betting}
\author{Rouven Michels\thanks{corresponding author: r.michels@uni-bielefeld.de} \thanks{ Bielefeld University, Universitätsstraße 25, 33615 Bielefeld, Germany}, Marius \"Otting\footnotemark[2], Roland Langrock\footnotemark[2]}

\date{}

\maketitle
\begin{abstract} 
\noindent
It is still largely unclear to what extent bettors update their prior assumptions about the strength and form of competing teams considering the dynamics during the match.  
This is of interest not only from the psychological perspective, but also as the pricing of live odds ideally should be driven both by the (objective) outcome probabilities and also the bettors' behaviour. 
Using state-space models (SSMs) to account for the dynamically evolving latent sentiment of the betting market, we analyse a unique high-frequency data set on stakes placed during the match.
We find that stakes in the live-betting market are driven both by perceived pre-game strength and by in-game strength, the latter as measured by the Valuing Actions by Estimating Probabilities (VAEP) approach.
Both effects vary over the course of the match. 
\end{abstract}
\keywords{OR in sports, live betting, state-space model, stochastic volatility, time series analysis}

\section{Introduction}

Given the economic relevance of betting markets --- the gross gaming revenue was reported as 41.7 billion Euro in 2020 in Europe \citep{europeanbetting} --- it is of much interest to understand the behaviour of the market's participants, i.e.\ the bettors.
From the bookmakers' perspective, it is first of all crucial to avoid inefficiencies in the pricing of odds. Furthermore, for profit maximisation it may be beneficial to exploit particular patterns in bettors' actions, such as a favourite-longshot bias (or the reverse bias). Such patterns and potential biases are also interesting from the psychological perspective, and several studies have indeed already focused on betting behaviour, e.g.\ investigating the `gambler's fallacy' \citep{tversky1974judgment, clotfelter1993gambler}, the reaction to a perceived `hot hand' \citep{sundali2006biases,paul2014bettor}, but also the drivers of demand for sports bets more generally \citep{humphreys2013consumption,deutscher2019demand}.

Sports betting takes place both in the pre-game (bets placed before kick-off) and in the in-game (bets placed during games) market. 
Although only 45\% of the betting volume in Europe (\citealp{europeanbetting}) is placed in the pre-game market, empirical research to date has largely focused on this market, investigating market (in-)efficiencies, over- and underreaction of bettors, drivers of betting volume and forecasting of match outcomes, to name but a few (see, e.g. \citealp{thaler1988anomalies,vergin2001overreaction, dixon2004value,feddersen2017sentiment,deutscher2018betting,BROWN20191073, butler2021expert, durand}). 
In contrast, for the in-game market, the existing literature to date has mostly focused on inefficiencies (see, e.g., \citealp{debnath2003information, choi2014role, croxson2014information}), such that in particular the betting behaviour in this important and highly dynamic market is not yet well understood. 
The principle question motivating this contribution thus is the following: What are the drivers of bet placements in the in-game market? And in particular: To what extent are these related to match dynamics?

For the pre-game betting market, building a forecast based on information such as the (perceived) teams' strength is comparable, conceptually, to a \textit{fundamental} analysis in the financial market \citep{abarbanell1997fundamental}. In the latter, the incentive to buy stocks --- the counterpart of a sports bet placement --- is that people feel confident that the `true' value of a company is larger than the actual stock price. Market analysts estimate the former by means of, e.g., the debt-equity-ratio and the earnings before interest, taxes, depreciation, and amortisation (EBITDA; \citealp{mukherji1997fundamental}, \citealp{quirin2000fundamental}, \citealp{baresa2013strategy}). Thus, investments based on a fundamental analysis correspond to an expectation of net profits in the long run. 
In contrast, exploiting short-time fluctuations in the financial market can be attempted based on a \textit{technical} analysis (see, e.g. \citealp{brown1989technical, edwards2018technical}).
In that case, the investor examines several quantitative indicators such as moving averages 
to build a short-term forecast \citep{zhu2009technical}.
While we do not want to stretch the analogy too much, we argue that in-game bettors can, in principle, follow not only a strategy similar to a fundamental analysis, essentially considering the current score relative to fixed variables such as the team strength, but also a strategy more similar to a technical analysis, incorporating information from short-term measures related for example to ball possession, tackles, goal-scoring opportunities or passes completed. To what extent either of these two strategy types drives the placement of bets in the in-game market is the focus of this work.

We investigate the effects of both fixed team information and in-game dynamics on the stakes placed in the in-game market based on two large and high-resolution data sets from the 2017/18 season of the German Bundesliga. The first data set covers detailed betting data, specifically in-game betting odds and volumes for all 306 matches played in the 2017/18 season. This unique data set was provided by a large European bookmaker and allows to investigate not only in-game odds (as has been done in the existing literature) but also stakes placed during the match. The second data set, provided by \citet{luca}, comprises WyScout event data, consisting of information on events such as shots on goal, passes, 1-on-1 situations, and set pieces for all 306 matches. The combination of these two large data sets enables us to investigate the potential effects of in-game dynamics on betting behaviour (rather than on the match outcome; such dependence has been studied, e.g., in \mbox{\citealp{weimar2017moneyball}, \citealp{brechot2020dealing}}). Moreover, we investigate how these effects may vary over the course of a match --- the expectation being that the importance of in-game (pre-game) information will increase (decrease) over time.

\section{Data}
We use in-game stakes placed during all 306 matches of the 2017/2018 German Bundesliga season to investigate the investment behaviour of bettors during football matches. In particular, the corresponding data considered cover bets on the match outcome, i.e.\ home win or away win (we exclude bets on a draw from our analysis). These data, which were provided to us by one of the largest bookmakers in Europe (with most of its customers located in Germany), have a 1 Hz resolution.
This temporal resolution is finer than necessary with respect to our research objective, such that to simplify the modelling we aggregate the second-by-second stakes into intervals of one minute. To compare stakes across teams and matches, we calculate \textit{relativestake} per team and minute, where for each interval we divide the amount of stakes placed on each team by the total amount placed on either team. 
The two teams' relative stakes thus always sum to $1$, such that it is sufficient to analyse the relative stakes from the point of view of one team only. Therefore, without loss of generality, in the following we will consider the relative stakes placed on the home team only.
The processed data set then comprises $N = 306$ time series, $\{ y_{n,t} \}$, $n=1,\ldots,306$, with $t$, $t = 1, \ldots, T$, indicating the minute of the match. Towards the end of the injury time, much less stakes are placed, such that we truncate all time series at minute $t=85$ to avoid the need to deal with the much increased variation. 
Overall, $26\,010$ observations of relative stakes are considered.

To investigate the drivers of betting behaviour, as represented here by the relative stakes being placed, we consider both static (pre-game) as well as dynamic (in-game) covariate information. For the former, we use the pre-game odds as a proxy for the (relative) pre-game strength of a team, as betting odds in general are accurate probabilistic forecasts for match outcomes (see, e.g., \citealp{spann2009sports}), and hence constitute reliable measures for the strength of teams (relative to their opponents). To simplify interpretation, we consider the implied probabilities, i.e. the inverse of the decimal odds, adjusted for the overround (or `vig').
To account for both teams' strength, we consider the difference of their pre-game winning probabilities (\textit{prewindiff}).

To additionally investigate the effect of in-game actions on the stakes placed, an additional comprehensive data set on in-game match events is considered. These data, again collected at 1 Hz, were provided by the company WyScout and were made publicly available by \citet{luca}. They contain information on all relevant actions during the match, together with a time stamp as well as the associated location on the pitch, indicated by $x$-- and $y$--coordinates. 
From the WyScout in-game data, we extract covariate information related to the match action, aiming to measure how the in-game team strength evolves dynamically throughout the match. 
To this end, several simple summary statistics could be used, e.g.\ the number of shots, the number of passes or the average distance of actions to the opposing goal. However, such simple metrics have been shown to be relatively poor predictors for the match outcome in football \citep{mackenzie,carling}. 

\begin{figure}[!h]
\centering
\includegraphics[scale = 1.25]{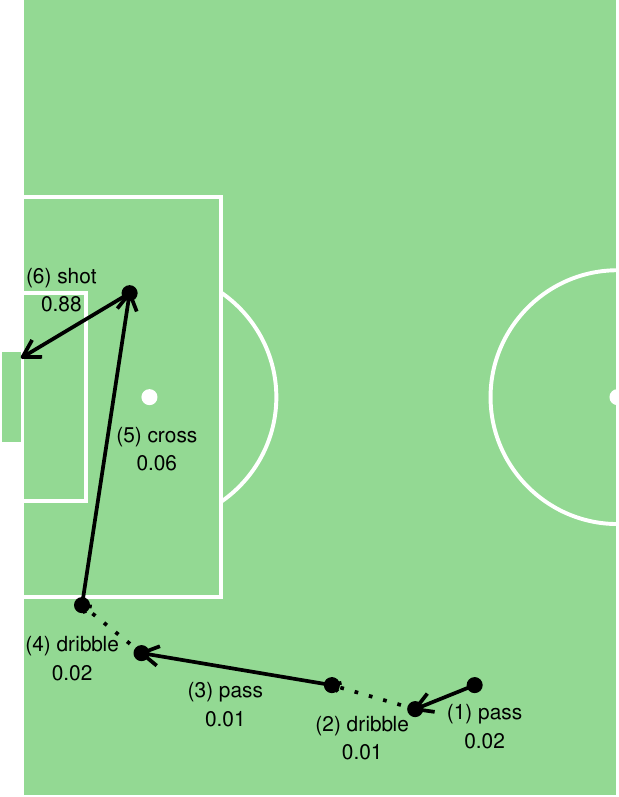}
\caption{\label{fig:label} The attack which led to a goal scored by Amine Harit in the match between Borussia Dortmund and FC Schalke 04 (4:4) on November 25, 2017. The numbers indicate the VAEP values of the corresponding actions.}
\end{figure}

\begin{figure}[!t]
	\centering
	\includegraphics[width=\textwidth]{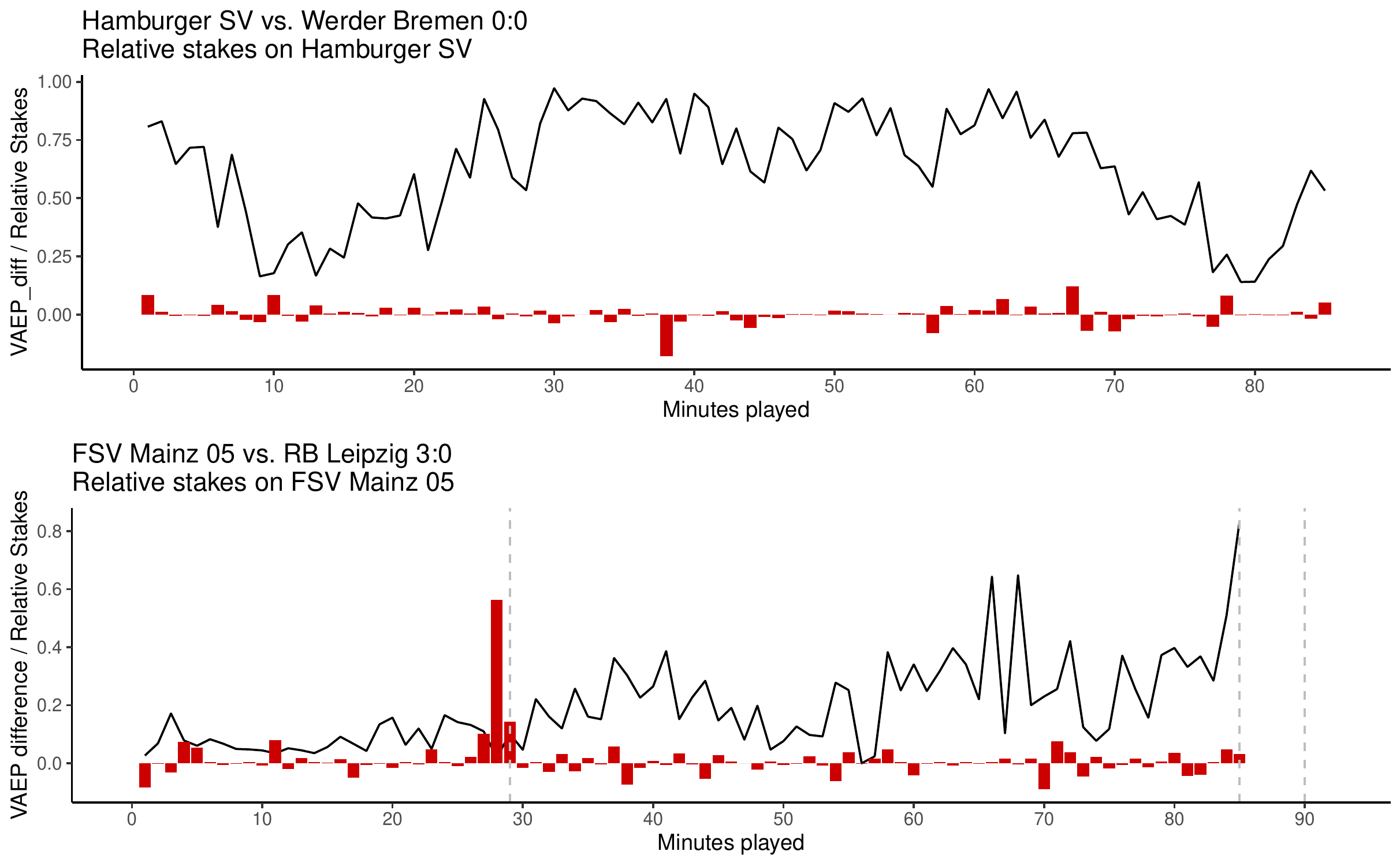}
	\caption{Time series of the relative stakes placed on a win of Hamburger SV (playing against Werder Bremen) and FSV Mainz 05 (playing against RB Leipzig). 
	The vertical bars denote the differences between the competing teams' VAEP values from the perspective of Hamburg and Mainz, respectively, and dashed lines indicate goals. In the bottom panel, the 2-0 and 3-0 were scored in minutes 85 and 90, respectively --- we truncate all time series in minute 85.}
	\label{img:grafik-dummy}
\end{figure}

To capture the in-game strength of a team appropriately we thus consider a more refined metric for measuring the \textit{value} of actions on the pitch, namely the so-called Valuing Actions by Estimating Probabilities (VAEP) approach \citep{vaep}. 
The idea of the VAEP is to measure the value of any action, e.g.\ a pass or a tackle, with respect to both the probability of scoring and the probability of conceding a goal. For illustration, Figure \ref{fig:label} shows an example sequence of actions and their associated VAEP values, obtained using predictive machine learning methods, in particular gradient-boosted trees --- see the Appendix for more details. From the action-level VAEP values, we build the covariate \textit{vaepdiff}, where we consider the differences between the teams' VAEP values aggregated over 1-minute intervals. The higher the value of \textit{vaepdiff}, the more the momentum of the match is with the team for which the relative stakes are modelled. 
The top panel in Figure \ref{img:grafik-dummy} shows an example of how \textit{vaepdiff} evolves over time along with our response variable, the relative stakes, for a scoreless draw. A sequence of positive VAEP differences between minutes 15 and 25 is followed by a shift towards increased betting on Hamburger SV, whereas around minute 40 a decrease in the relative stakes may be caused by negative VAEP differences, i.e. Werder Bremen gaining some momentum. The bottom panel of Figure \ref{img:grafik-dummy} shows a second example match, one in which goals were scored. The first goal in minute 29 here leads to a shift towards bets on FSV Mainz 05. However, the relative stakes are mostly smaller than 0.5, likely caused by RB Leipzig being the favourite in the match, as their pre-game winning probability was larger than for FSV Mainz 05.

\begin{figure}[!t]
	\centering
	\includegraphics[width=\textwidth]{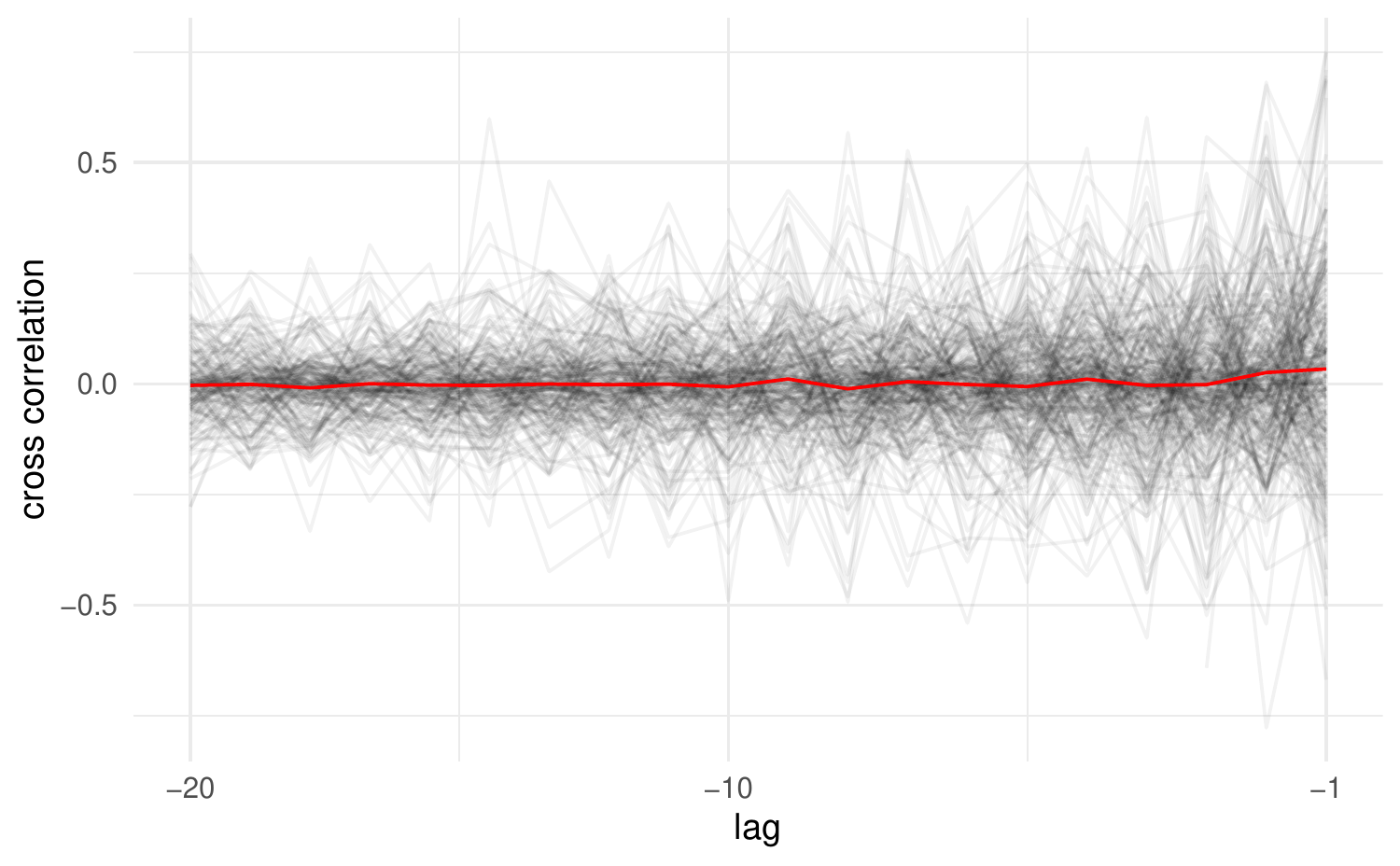}
	\caption{Cross-correlations between the VAEP and the relative stakes, with each grey line referring to one of the 306 time series considered, and the red line showing the overall mean for each lag.}
	\label{img:ccfs}
\end{figure}

To explore the relationship between the VAEP and the relative stakes in our data, we consider their serial cross-correlations. Figure \ref{img:ccfs} shows the first 20 cross-correlation lags for all matches in our data (truncated at the time the first goal was scored, to allow for a meaningful comparison), indicating a small positive correlation at lower lags.
Table \ref{all} displays summary statistics on the VAEP and on all remaining variables considered in our analysis.

\begin{table}[!h] \centering 
  \caption{Descriptive statistics of the variable analysed,
  \textit{relativestake}, as well as the covariates \textit{prewindiff} and \textit{vaepdiff}. Since the covariate \textit{prewindiff} is constant across matches, the standard deviation was calculated on the basis of one value per match.}
  \label{all} 
\begin{tabular}{@{\extracolsep{5pt}}lccccccc} 
\\[-1.8ex]\hline 
\hline \\[-1.8ex] 
 & \multicolumn{1}{c}{n} & \multicolumn{1}{c}{mean} & \multicolumn{1}{c}{st.\ dev.} & \multicolumn{1}{c}{min} & \multicolumn{1}{c}{max} \\ 
\hline \\[-1.8ex] 
\textit{relativestake} & 26,010 & 0.493 & 0.313 & 0.000 & 1.000 \\ 
\textit{prewindiff} & 26,010 & 0.139 & 0.317 & $-$0.740 & 0.851 \\ 
\textit{vaepdiff} & 26,010 & 0.004 & 0.161 & $-$1.091 & 1.167 \\ 
\hline \\[-1.8ex] 
\end{tabular} 
\end{table}

\section{Modelling relative stakes}\label{chap:models}

The time series of relative stakes shows strong positive serial correlation (cf.\ Figure \ref{img:grafik-dummy}). This correlation does not result from a potential direct effect of the stakes placed in any given minute on the stakes placed in subsequent one-minute intervals, as individual bettors (for the most part) act independently of each other. Instead, the correlation is induced by the market progressing through different phases: for example, in the top panel of Figure \ref{img:grafik-dummy}, the relative stakes placed on the home team are relatively low between minutes 10--15. In contrast, between minutes 25 and 65, bets are placed predominantly on the home team.
Such different phases can be formalised as a latent variable within a state-space model (SSM), which therefore constitutes the natural modelling approach for our data: stakes are driven by the current underlying market phase, more intuitively to be understood as the market sentiment (e.g. leaning towards bets being placed on the home team), and the unobserved market sentiment evolves probabilistically over time, exhibiting serial correlation and hence inducing serial correlation also in the observed time series of stakes. 

More specifically, such an SSM comprises two processes, (i) an unobserved, serially correlated state process $\{ g_{n,t} \}$, in our application to be seen as a proxy of the market sentiment, with higher values implying a higher tendency towards bet placement on the home team, 
and (ii) an observed state-dependent process $\{ y_{n,t} \}$, which is driven by $\{ g_{n,t} \}$ and in our case is the time series of relative stakes placed. For simplicity, the subscript $n$, indicating the match considered, is omitted in the following. The state variables are assumed to be first-order Markovian, i.e.\ 
\begin{linenomath*}
\begin{equation*}
f(g_{t} | g_1,\ldots,g_{t-1}) = f(g_{t} | g_{t-1}),
\end{equation*}
\end{linenomath*}
and the observations are assumed to be conditionally independent of each other and of previous states, given the current state:
\begin{linenomath*}
\begin{equation*}
f(y_t | g_1,\ldots,g_t, y_1,\ldots,y_{t-1}) = f(y_t | g_t). 
\end{equation*}
\end{linenomath*}
In the next section, we develop the specific model formulation, i.e.\ the precise form of the conditional distributions $f(g_{t} | g_{t-1})$ and $f(y_t | g_t)$, used to model the time series of relative stakes. Details on the implementation of the maximum likelihood estimation of the model parameters are provided in the Appendix. 

\subsection{Baseline state-space model}

\begin{figure}[!h]
    \centering
	\begin{tikzpicture}
	\node[circle,draw=black, fill=gray!5, inner sep=0pt, minimum size=50pt, text width=1.3cm] (COV1) at (2, 0) {\textit{prewin-}\textit{diff}};
	\node[circle,draw=black, fill=gray!5, inner sep=0pt, minimum size=50pt, text width=1.3cm] (COV2) at (4.5, 0) {\textit{prewin-}\textit{diff}};
	\node[circle,draw=black, fill=gray!5, inner sep=0pt, minimum size=50pt, text width=1.3cm] (COV3) at (7, 0) {\textit{prewin-}\textit{diff}};
	\node[circle,draw=black, fill=gray!5, inner sep=0pt, minimum size=50pt, text width=1.3cm] (COV4) at (2, -7.5) {\textit{vaep}- \textit{diff}$_{t-2}$};
	\node[circle,draw=black, fill=gray!5, inner sep=0pt, minimum size=50pt, text width=1.3cm] (COV5) at (4.5, -7.5) {\textit{vaep}- \textit{diff}$_{t-1}$};
	\node[circle,draw=black, fill=gray!5, inner sep=0pt, minimum size=50pt, text width=1.3cm] (COV6) at (7, -7.5) {\textit{vaep}- \textit{diff}$_{t}$};
	\node[circle,draw=black, fill=gray!5, inner sep=0pt, minimum size=50pt] (A) at (2, -5) {$g_{t-1}$};
	\node[circle,draw=black, fill=gray!5, inner sep=0pt, minimum size=50pt] (A1) at (-0.5, -5) {...};
	\node[circle,draw=black, fill=gray!5, inner sep=0pt, minimum size=50pt] (B) at (4.5, -5) {$g_{t}$};
	\node[circle,draw=black, fill=gray!5, inner sep=0pt, minimum size=50pt] (C) at (7, -5) {$g_{t+1}$};
	\node[circle,draw=black, fill=gray!5, inner sep=0pt, minimum size=50pt] (C1) at (9.5, -5) {...};
	\node[circle,draw=black, fill=gray!5, inner sep=0pt, minimum size=50pt] (Y1) at (2, -2.5) {$y_{t-1}$};
	\node[circle,draw=black, fill=gray!5, inner sep=0pt, minimum size=50pt] (Y2) at (4.5, -2.5) {$y_{t}$};
	\node[circle,draw=black, fill=gray!5, inner sep=0pt, minimum size=50pt] (Y3) at (7, -2.5) {$y_{t+1}$};
	\draw[-{Latex[scale=2]}] (A)--(B);
	\draw[-{Latex[scale=2]}] (B)--(C);
	\draw[-{Latex[scale=2]}] (A1)--(A);
	\draw[-{Latex[scale=2]}] (C)--(C1);
	\draw[-{Latex[scale=2]}] (A)--(Y1);
	\draw[-{Latex[scale=2]}] (B)--(Y2);
	\draw[-{Latex[scale=2]}] (C)--(Y3);
	\draw[-{Latex[scale=2]}] (COV1)--(Y1);
	\draw[-{Latex[scale=2]}] (COV2)--(Y2);
	\draw[-{Latex[scale=2]}] (COV3)--(Y3);
	\draw[-{Latex[scale=2]}] (COV4)--(A);
	\draw[-{Latex[scale=2]}] (COV5)--(B);
	\draw[-{Latex[scale=2]}] (COV6)--(C);
	\end{tikzpicture}
\caption{Dependence structure of the SSM used to model the relative stakes $y_t$, driven by the state variable $g_{t}$ corresponding to the market sentiment. Additional covariate dependence is assumed in both state-dependent and state process, for the former considering static covariate information (pre-game winning probabilities are constant within matches), for the latter considering dynamic covariate information built from event data.}
\label{fig:HMM}
\end{figure}
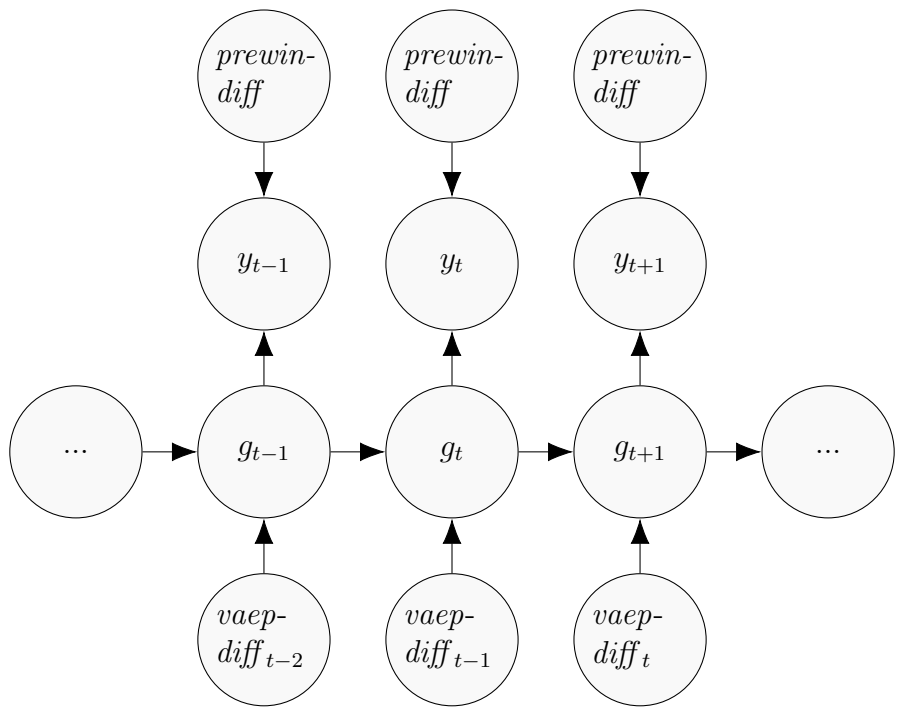

Our response variable $y_t$, the relative amount of stakes placed on the home team, is continuous-valued with support $[0,1]$, rendering the beta distribution a natural choice for modelling purposes. The support of the regular beta distribution is $(0,1)$, such that we use the beta-inflated distribution (BEINF) to account for the fact that in some intervals stakes are placed on one team only (in which case $y_t=0$ or $y_t=1$). 
We follow the parametrisation proposed by \citet{rigby2019distributions}, such that
\begin{linenomath*}
$$
y_t \sim \text{BEINF}(\mu_t, \sigma, p, q), \quad \text{with} \quad
f(y_t) = \begin{cases} p, \text{ for } y_t=0; \\ (1 - p - q) h(y_t), \text{ for } 0<y_t<1; \\ q , \text{ for } y_t=1 , \end{cases}
$$
for $0 \leq y_t \leq 1$. Here $h(y_t)$ is the density function of the regular beta distribution, i.e.
$$
h(y_t, a, b) = \frac{y_t^{a-1}(1-y_t)^{b - 1}}{\text{B}(a, b)},
$$
\end{linenomath*}
with the beta function $\text{B}(a,b)$. The shape parameters $a$ and $b$ are not directly amenable to regression modelling, such that we consider a reparametrisation in terms of the beta distribution's mean $\mu_t$ and its standard deviation $\sigma$ \citep{rigby2019distributions}. From $\mu_t$ and $\sigma$ the associated shape parameters are obtained as $a = \mu_t(1-\sigma^2)/\sigma^2$ and $b = (1-\mu_t)(1-\sigma^2)/\sigma^2$. Figure \ref{fig:HMM} displays the exact dependence structure of our SSM as a directed acyclic graph. Specifically, to account for the dynamic nature of the relative stakes within matches, the mean $\mu_t$ is assumed to be time-varying and is modelled as follows:
\begin{equation}\label{eq1}
\mu_t = \text{logit}^{-1} (\alpha_0 + \alpha  \textit{prewindiff} + g_t).
\end{equation}
The rationale of the latter two components of the linear predictor is as follows.  
We include the difference of the teams' pre-game win probabilities (\textit{prewindiff}) to address systematic effects such as potential favourite-longshot biases, or their reverse, throughout matches (see, e.g., \citealp{cain2000favourite}, \citealp{cain2003favourite}).
In addition, information on the current market sentiment is included via $g_t$, thereby accounting for phases with stronger 
preferences in bet placement on either team. If a team shows positive actions on the pitch, then the market sentiment towards placing bets on that team should improve, which would be reflected by an increase in $g_t$.

The unobserved variable corresponding to the market sentiment, $g_t$, is modelled as an autoregressive process of order 1, with additional covariate dependence:
\begin{equation}\label{eq2}
g_t = \phi g_{t-1} + \beta \textit{vaepdiff}_{t-1} + \omega \eta_t,
\end{equation}
with $\eta_t \overset{\text{iid}}{\sim} \mathcal{N}(0,1)$, $\omega > 0$ and $\phi \in (-1,1)$. The in-game covariate $\textit{vaepdiff}$ is included in the state variable $g_t$, as it seems natural to assume that potential effects of positive actions by a team will not necessarily be instantaneous --- i.e.\ affecting the mean of the relative stakes only in the very minute the action took place --- but rather accumulate and persist over some time. For example, from the bettor's perspective, a single one-minute interval with high $\textit{vaepdiff}$ values is not as likely to affect his or her betting decision as a positive spell of say 15 minutes of the team considered, with overall elevated $\textit{vaepdiff}$ values. This is accounted for by allowing positive contributions of $\textit{vaepdiff}$ to accumulate in $g_t$, which for $\phi>0$ is persistent and hence to some extent memorises these contributions.

\subsection{Varying-coefficient state-space model}

The model described in the previous section is to be regarded as a baseline model, which attempts to formalise a) how the relative stakes evolve during a match depending on the market sentiment, and b) how the covariates of primary interest, i.e. \textit{prewindiff} and \textit{vaepdiff}, are most naturally incorporated into this model.
We will now extend this baseline model to improve its realism by allowing the effects of these covariates to vary over time and by introducing additional control variables.

\subsubsection*{Varying coefficients}

Due to the dynamic nature of football matches, it may very well be the case that the effects of \textit{prewindiff} and \textit{vaepdiff} vary over time. Specifically, it seems intuitively plausible that the strengths of the teams, as measured by the pre-game win probabilities, are a strong predictor for betting activity at the beginning of a match when little additional information is available. Vice versa, towards the end of a match the teams' performances during the match might become increasingly important for understanding betting dynamics. 
To allow the effects of \textit{prewindiff} and \textit{vaepdiff} to vary over time, we replace $\alpha$ and $\beta$ in Eqs.\ (\ref{eq1}) and (\ref{eq2}) by time-varying parameters $\alpha_t$ and $\beta_t$, respectively. To avoid a priori assumptions on the functional forms of $\alpha_t$ and $\beta_t$, we model these functions nonparametrically using B-splines.

Since the seminal paper by \citet{splines}, the class of B-splines has rapidly gained popularity in nonparametric statistical modelling, and in recent years, B-spline-based modelling of functional effects has been embedded also in various types of SSMs (see, e.g. \citealp{de2014switching,hambuckers2018markov,mews2022}). 
In our setting, $\alpha_t$ and $\beta_t$, the time-varying effects of \textit{prewindiff} (on the mean relative stakes) and \textit{vaepdiff} (on the state variable indicating the market sentiment), respectively, are modelled as linear combinations of a finite number of section-wise defined basis functions,
\begin{equation}\label{lincom}
      \alpha_t = \sum_{k = 1}^{K} \nu_{k}^{\alpha} B_k(t), \quad
      \beta_t = \sum_{k = 1}^{K} \nu_{k}^{\beta} B_k(t),
\end{equation}
for $t = 1, \ldots, 85$, where $B_1, \ldots, B_k$, $k = 1, \ldots, K$, are fixed, equidistant B-spline basis functions of order three. We use cubic polynomial B-splines to obtain a twice continuously differentiable function, thus leading to smooth density estimates \citep{langrock2017markov}.
To prevent overfitting we add a roughness penalty term, thus considering so-called penalised B-splines, i.e.\ P-splines \citep{splines}. Specifically, we penalise high values of the second-order differences of adjacent coefficients in the linear combinations above. The sum of these second-order differences corresponds to an approximation of the integrated squared curvature of the functional estimate. The resulting penalised log-likelihood function is then given as follows (cf.\ \citealp{langrock2017markov}):
\begin{linenomath*}
\begin{align*}
    \ell_p = \log \big (\mathcal{L}_{\text{approx}}\big ) -  \frac{\lambda_{\alpha}}{2} \sum_{k = 3}^K (\Delta^2 \nu_{k}^{\alpha})^2 -  \frac{\lambda_{\beta}}{2} \sum_{k = 3}^K (\Delta^2 \nu_{k}^{\beta})^2, 
\end{align*}
\end{linenomath*}
with the unpenalized likelihood function $\mathcal{L}_{\text{approx}}$ (see Eq.\ (\ref{unpen}) in the Appendix), the second-order differences $\Delta^2 \nu_{k} = \nu_{k}- 2 \nu_{k-1}+\nu_{k-2}$, and smoothing parameters $\lambda_{\alpha}$ and $\lambda_{\beta}$ to control the bias-variance trade-off. For $\lambda_{\alpha}, \lambda_{\beta} \rightarrow \infty$ the varying coefficients $\alpha_t$ and $\beta_t$ simplify to a linear effect (\citealp{splines}). In other words, this nonparametric approach can capture complex time-varying effects if necessary and otherwise will typically collapse to simple linear modelling due to the penalisation of non-zero curvature.

Following \citet{epub31413}, the tuning parameters $\lambda_{\alpha}$ and $\lambda_{\beta}$ are chosen via the Akaike Information Criterion (AIC), $\text{AIC} = -2 \cdot l + 2 \cdot \widehat{df}$, where $l$ is the unpenalised likelihood under the fitted model and $\widehat{df}$ is an estimate of the degrees of freedom. The latter is obtained as the trace of the product of the Fisher information matrix for the unpenalized likelihood and the inverse Fisher information matrix for the penalized likelihood (see \citealp{degrees}). We consider the following two-dimensional grid from which the smoothing parameters are chosen: 
\begin{equation}\label{tune}
      \Lambda_{\alpha} \times \Lambda_{\beta} = \{0.05, 0.25, 1, 5, 25, 100, 500\} \times \{0.05, 0.25, 1, 5, 25, 100, 500\}.
\end{equation}
For each combination of $\lambda_{\alpha} \in \Lambda_{\alpha}$ and $\lambda_{\beta} \in \Lambda_{\beta}$, the model is fitted and the AIC calculated, then selecting the combination of $\lambda_{\alpha}$ and $\lambda_{\beta}$ that yields the lowest AIC value. 
A more detailed discussion of the implementation of P-splines can be found in \citet{langrock2017markov}.

\subsubsection*{Additional control variables}

As indicated by the bottom panel of Figure \ref{img:grafik-dummy}, the relative stakes placed on the home team might also be affected by the current score. We thus now also include the difference in the current score at minute $t$ ($\textit{scorediff}_t$), calculated from the home team's point of view, such that positive values correspond to a lead of the home team. Since considering the difference in the current score alone does not fully reflect a team's winning chances, we further include the in-game win probability in minute $t$, which is derived by the betting odds in minute $t$ ($\textit{winprobteam}_t$). As for the pre-game win probabilities, we again consider the inverse of the odds adjusted for the bookmaker's vig. 

We add these further covariates to the predictor for the mean of the BEINF distribution:
\begin{equation}\label{eq:splines}
\mu_t = \text{logit}^{-1} \bigl(\alpha_0 + \sum_{k = 1}^{K} \nu_{k}^{\alpha} B_k(t) \ \textit{prewindiff} + \zeta_1 \textit{scorediff}_t + \zeta_2 \textit{winprobteam}_t + g_t\bigr).
\end{equation}

\section{Results}
\subsection{Baseline state-space model}\label{4.1}

For the baseline SSM specified by (\ref{eq1}) and (\ref{eq2}), with the effects of \textit{prewindiff} and \textit{vaepdiff} assumed to be constant over time, the parameter estimates are given in
Table \ref{tab:resultscovariates_state2}. 
The persistence in the state process was estimated to be fairly strong ($\hat{\phi} = 0.968$), corresponding to a positive correlation in the proportional allocation of stakes. In other words, if bets are placed predominantly on either of the two teams, this pattern tends to persist for some time. The estimate of the intercept $\alpha_0$ is negative, indicating that overall higher stakes are placed on the away team (see also \citealp{levitt}, for similar results) when the pre-game win probabilities of both teams are identical (i.e. $\textit{prewindiff} = 0$). A possible explanation of this effect is that bettors may be underestimating the home advantage.
The effects of \textit{prewindiff} and \textit{vaepdiff} were both estimated to be positive, confirming the intuition that team strength, both prior to the match and as manifested during the match itself, is valued by bettors. According to the AIC, this model is preferred over the model excluding \textit{prewindiff} ($\Delta$AIC = 339.17)  as well as over the model excluding \textit{vaepdiff} ($\Delta$AIC = 522.04). This is corroborated also by the 95\% confidence intervals given in Table \ref{tab:resultscovariates_state2}. 

\begin{table}[htb]
\centering
\caption{Parameter estimates with 95\% confidence intervals for the baseline SSM. 
}\vspace{0.3em}
\label{tab:resultscovariates_state2}
\begin{tabular}{lccc}
  \hline
 parameter & estimate & \multicolumn{1}{c}{95\% CI} \\ 
  \hline
$\phi$ & 0.968 & [0.964;\,0.971] \\ 
$\omega$ & 0.249 & [0.238;\,0.261] \\
$\sigma$ & 0.300 & [0.296;\,0.303] \\
\hdashline
$\alpha_0$ & -0.195 & [-0.278;\,-0.113] \\
$\alpha$ $\big(\textit{prewindiff}\big)$ & 2.395 & [2.151;\,2.640] \\
$\beta$ $\big(\textit{vaepdiff}\big)$ & 0.600 & [0.550;\,0.651]\\
   \hline
\end{tabular}
\end{table}

\subsection{Varying-coefficient state-space model}

\begin{figure}[!h]
	\centering
	\includegraphics[width=\textwidth]{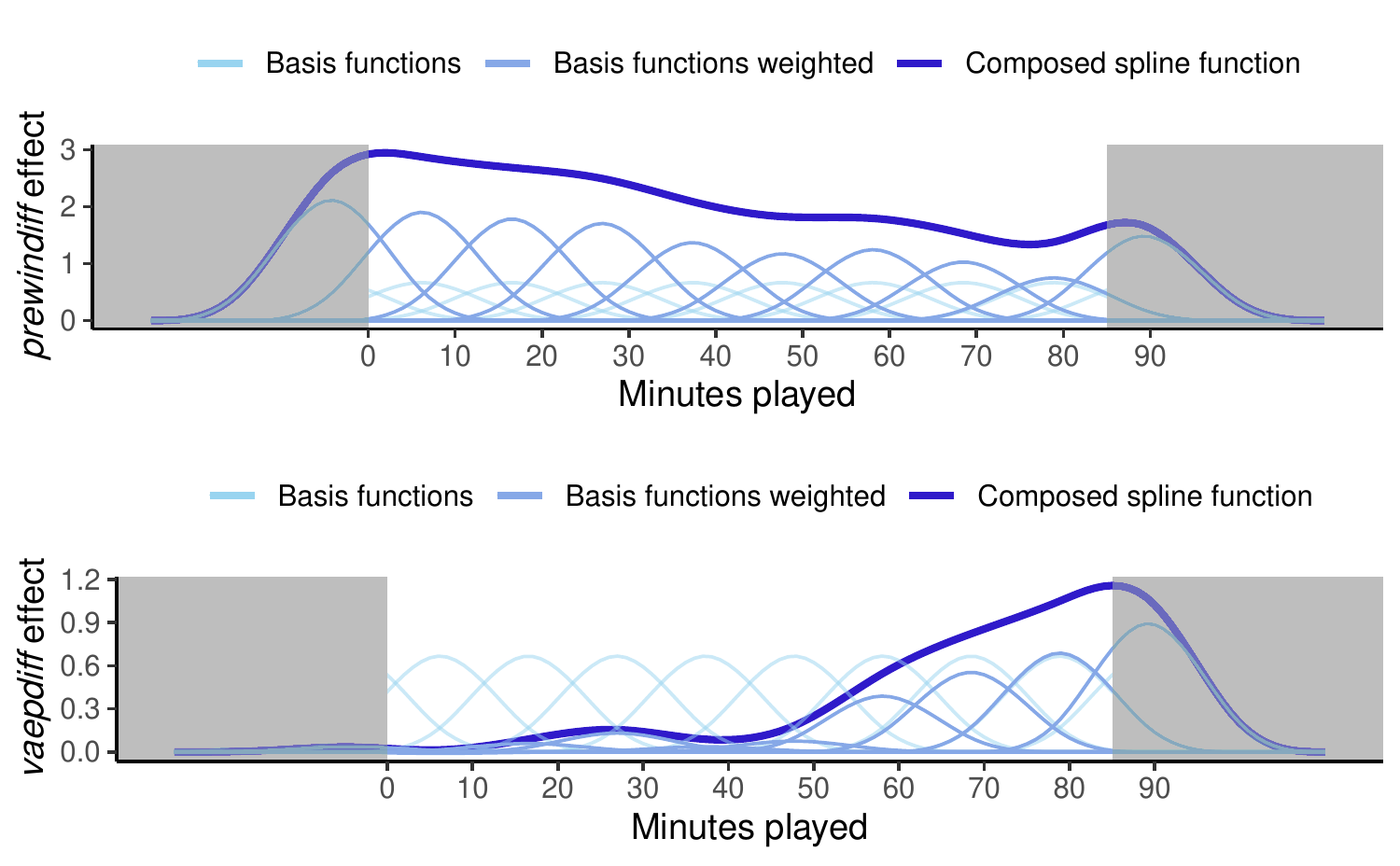}
	\caption{Time varying P-spline effects of the pre-game winning probabilities ($\lambda_{\alpha} = 1$) as well as the VAEP variable ($\lambda_{\beta} = 5)$. The white area indicates the intervals data was considered for. The grey area indicates where outer knots are set.}
	\label{splines}
\end{figure}

The varying-coefficient SSM, which allows the effects of \textit{prewindiff} and \textit{vaepdiff} to change as the match progresses, was estimated using $K = 10$ basis functions to build the functional effects according to (\ref{lincom}). The model was fitted for all combinations of tuning parameters from $\Lambda_{\alpha} \times \Lambda_{\beta}$ as specified in (\ref{tune}), with the optimal choice $(\lambda_{\alpha},\lambda_{\beta}) = (1,5)$ according to the AIC. 
The model including time-varying effects of \textit{prewindiff} and \textit{vaepdiff} is clearly favoured over the simpler model reported in Section \ref{4.1} ($\Delta$AIC = 270.62)\footnote{To select the tuning parameters $\lambda_{\alpha}$ and $\lambda_{\beta}$, we also considered the BIC and the Hannan-Quinn criterion, which led to very similar results.}.

Regarding the estimated parametric effects, the results again confirm serial correlation in the state process ($\hat\phi = 0.963$). When the market sentiment towards the home team is improved (as induced for example by a large \textit{vaepdiff} value), such that $g_t$ takes higher values, this process tends to persist in this phase for the next few minutes. 
For the effect of the difference in the current score and the in-game win probability, $\zeta_1$ and $\zeta_2$ are both estimated to be positive ($\hat{\zeta}_1 = 0.234,\ \hat{\zeta}_2 = 0.336$) --- teams having the lead and those with a higher in-game win probability are thus preferred by bettors.

The nonparametrically estimated time-varying effects of \textit{prewindiff} and \textit{vaepdiff} are shown in 
Figure \ref{splines}. The effect of \textit{prewindiff} on stake placement is estimated to be positive throughout the match, but with the effect size decreasing as the match progresses. This matches the intuition that the effect of \textit{prewindiff} --- a variable measuring the overall strength of a team but not taking into account the actions on the pitch --- should be largest when new information is limited, i.e.\ at the very beginning of matches. The estimated effect size of \textit{prewindiff} decreases approximately linearly over time. 
In contrast, for \textit{vaepdiff} we find a highly non-linear functional form of the effect size over time. The effect of \textit{vaepdiff} --- a variable measuring the in-game strength of a team --- is estimated to be slightly positive throughout the first half of a match, followed by a much more rapid increase in the second half. Bettors thus do value actions on the pitch, with positive spells of a team leading to a shift in the market's sentiment and eventually an increase in the relative stakes placed on that team. Perhaps most interestingly, the importance of in-game actions as drivers of betting volumes very rapidly increases towards the end of matches. 

The results show that bettors incorporate information both on the perceived (pre-game) team quality as well as in-game dynamics when devising their betting strategy. This is in accordance with evidence found for financial markets, where more than 85\% of investors rely on both fundamental as well as technical analysis \citep{lui1998use}, thus incorporating both the valuation of a company as well as the stock's more short-term momentum.  

\subsection{Implications for bookmakers}

\begin{figure}[!h]
	\centering
	\includegraphics[scale = 0.9]{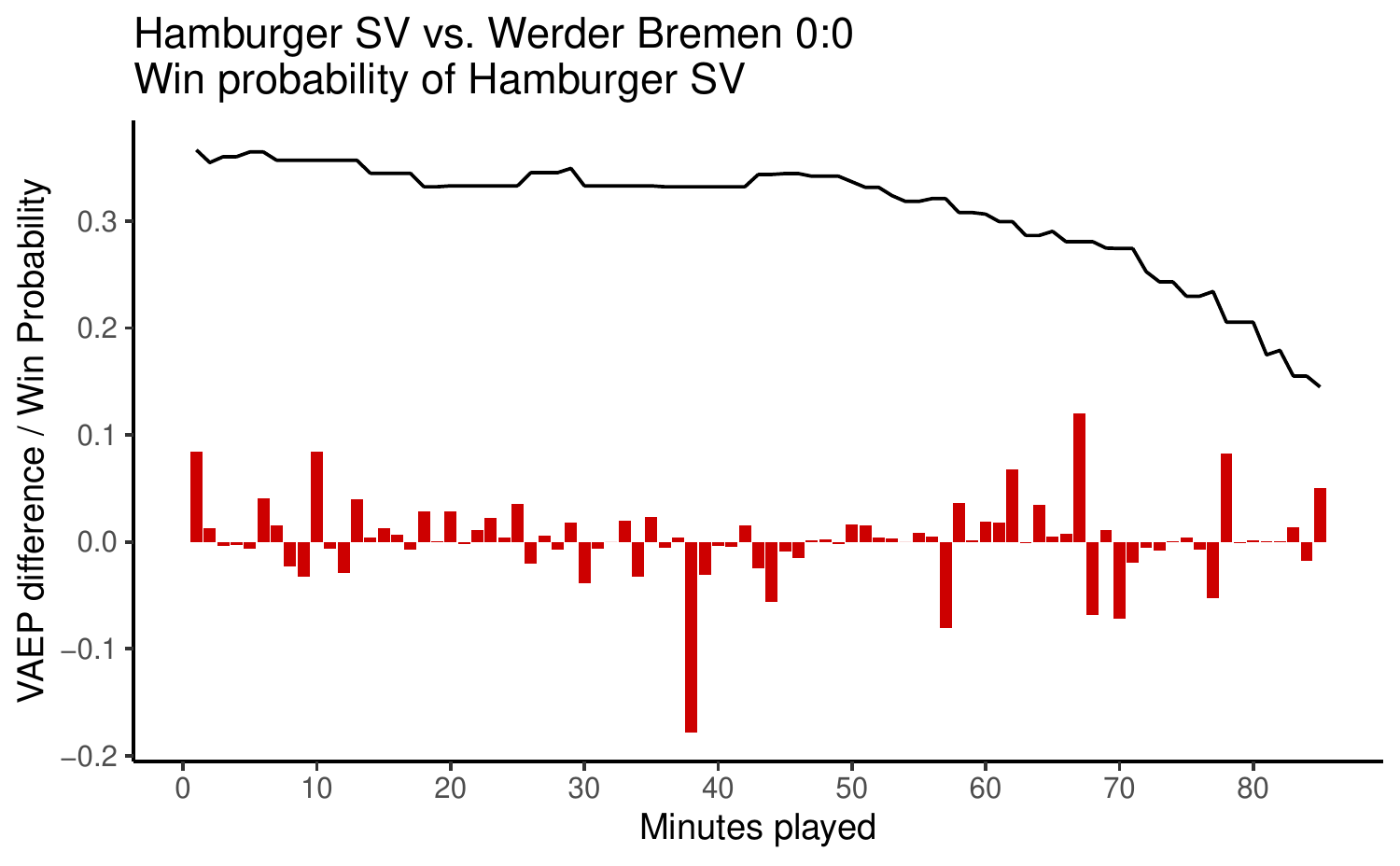}
	\caption{Time series of the win probabilities of Hamburger SV for one example match from the data set (Hamburger SV vs. Werder Bremen). The vertical bars denote the differences between both teams' VAEP values from the point of view of Hamburger SV. 
	Despite phases of dominance seeming to alternate, especially towards the end of the match between minutes 60 and 80, the corresponding win probability --- derived by the in-game odds --- does not fluctuate accordingly. The changes in the win probability are almost exclusively resulting from the time remaining.}
	\label{winprobs}
\end{figure}

The estimated effect of in-game actions on stake placement indicates that bettors' stake placement is driven by in-game dynamics. In contrast, for the example match Hamburger SV vs.\ Werder Bremen, Figure \ref{winprobs} illustrates that the betting odds and hence the implied winning probabilities appear to 
be largely unaffected by in-game actions as proxied by the VAEP differences. As bookmakers thus do not seem to incorporate such information into their betting odds, this raises the question of whether bettors can exploit such potential inefficiencies. To that end, we evaluate an example simple betting strategy: we consider all matches that are tied after halftime and evaluate the outcome of a strategy placing a bet of 1 euro on the home team whenever the \textit{vaepdiff} exceeds certain thresholds.
Table \ref{tab:returns} displays the returns when following this strategy for our data set, distinguishing several match periods. 
While positive returns could have been obtained for relatively large \textit{vaepdiff} values in the final phase of a match, it should be noted here that this combination involves only a few data points. 
The betting strategy would in fact have led to negative returns in the majority of settings, with some of the negative returns substantially larger than the usual vig of about 5\%.

\begin{table}[ht]
\centering
\caption{Possible returns for bettors using simple in-game betting strategies on final match outcomes based on different thresholds for \textit{vaepdiff}. Only matches tied in the second half are considered.}
\label{tab:returns}
\scalebox{0.72}{
\begin{tabular}{rcccccc}
  \hline
 & \textit{vaepdiff}$>0.02$ & \textit{vaepdiff}$>0.03$ & \textit{vaepdiff}$>0.05$ \\ 
  \hline
minute 45-60 
& -0.27 & -0.20 & -0.14 \\ 
minute 60-75 
& -0.08 & -0.17 & -0.28 \\ 
minute 75 until end 
& -0.01 & 0.04 & 0.32 \\ 
   \hline
\end{tabular}}
\end{table}

The models developed here could also be used by bookmakers, for example when setting odds or developing automatic fraud detection systems. In particular, the models can be used to study the sensitivity of stake placement with respect to changes in the odds, thereby potentially identifying opportunities for profit maximisation. Furthermore, model-based forecasts of relative stakes can be used to inspect unusual betting behaviour, i.e. to identify stake placements that are not well-explained by the model.

\begin{figure}[!h]
	\centering
	\includegraphics[scale = 0.8]{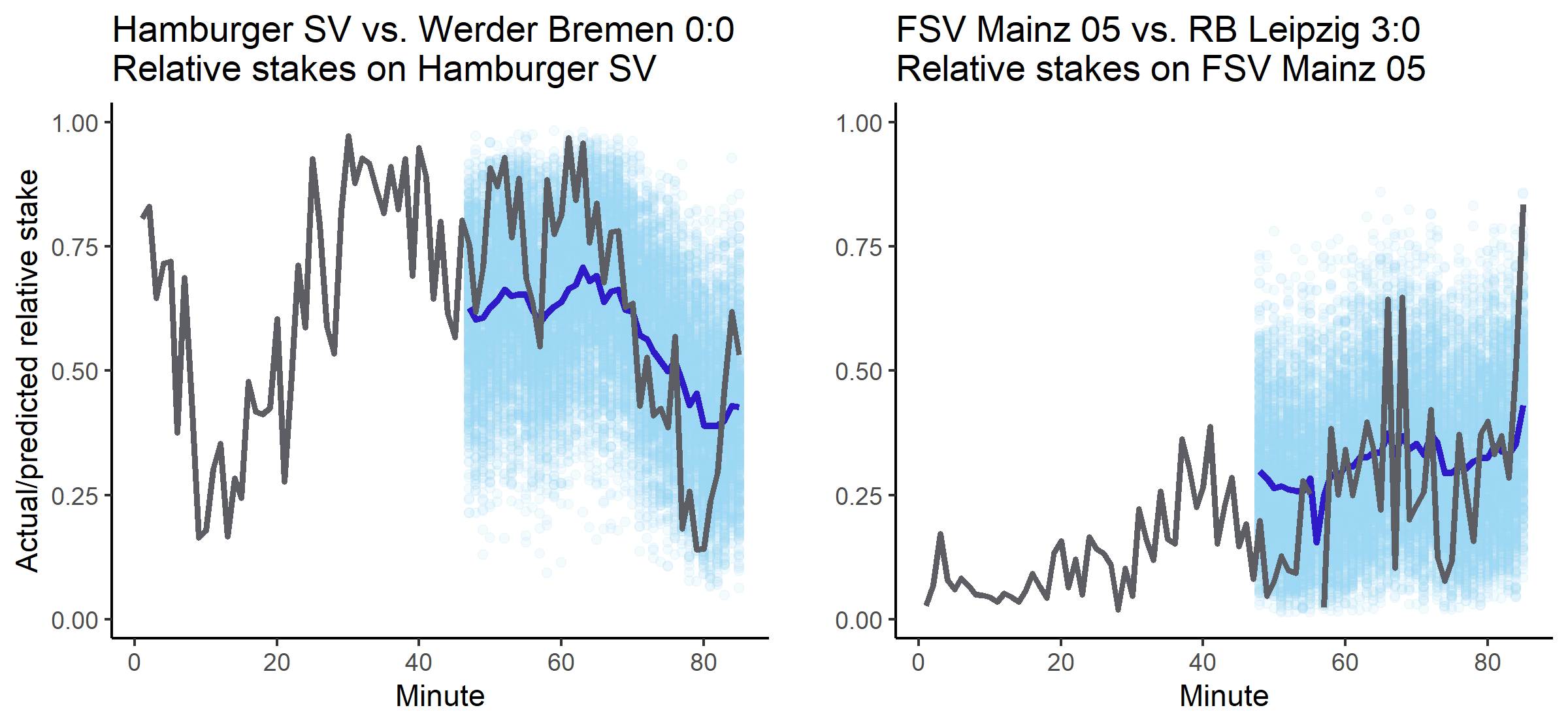}
	\caption{Observed relative stakes (grey line) and one-step-ahead predictions of relative stakes under our model and conditional on the first half. To illustrate the uncertainty, the light blue points are simulated from the one-step-ahead predicted BEINF distribution at time $t+1$ (500 simulated observations for each time point). The blue line indicates the predicted mean.}
	\label{fig:predict}
\end{figure}

To obtain real-time predictions of relative stakes, depending on the odds, we can calculate one-step-ahead forecasts under the fitted SSM. 
For this, we use the same discretisation strategy as for the parameter estimation (cf. Appendix), such that the state distribution in minute $t+1$ is predicted, based on which we obtain the distribution of relative stakes according to Eq. (\ref{eq:splines}). 
To illustrate such one-step-ahead prediction, Figure \ref{fig:predict} displays the time series of relative stakes placed within the example matches already shown in Figure \ref{img:grafik-dummy}, together with the model-based one-step-ahead forecasts during the second half. In addition to showcasing how the model can be used in particular for outlier detection, the figure also indicates that our model provides adequate predictions of the relative stakes for the two example time series.

\section{Discussion}

Our results show that bettors react to in-game dynamics. With this effect being strongest in the second half, it seems that bettors try to exploit the information provided by in-game dynamics. However, evaluating a corresponding simple betting strategy where bets are placed on teams with positive actions, we found that the potential returns to bettors are negative --- thus indicating an overreaction by bettors. Such an overreaction to positive events has in fact already been reported in sports betting (\citealp{durand,otting2021reaction}) and is also known to occur in financial markets (see, e.g., \citealp{tang,extrem}). In particular, \citet{huang2007expected} showed that price-earning ratios in the financial market deviate from expectations based on fundamentals, in other words that stock prices can be subject to an overreaction. For the in-game betting market, our results indicate a similar momentum-induced overreaction by bettors.

From a bookmaker's point of view, 
the overreaction of bettors described above can improve profits. Our analysis does however not reveal whether the bookmaker's current pricing strategy is profit-maximising.
A framework for maximising bookmaker revenue by taking into account in-game information was developed by \citet{lorig2021optimal}.

Moreover, for bookmakers, the model-based prediction of relative stakes can be beneficial in detecting unusual betting behaviour. In particular, if the observed relative stakes are substantially higher than the predicted relative stakes --- e.g. outside the 99\% quantile of the associated forecast distribution --- this could indicate insider information or fraud. However, a comprehensive fraud detection system should additionally take into account the absolute (rather than only the relative) stakes \citep{otting2018integrating}.

In more general terms, while our study focused on a specific aspect of live betting --- namely to what extent in-game dynamics affect bettor behaviour --- it also illustrates the immense potential of sports data, the availability of which has improved thanks to data providers such as WyScout or StatsBomb. The analysis of such complex data requires sophisticated statistical modelling, and SSMs as applied in the present paper constitute a versatile framework to accommodate the time series nature of most sports data (cf.\ \citealp{koopmeiners2012comparison,green2018hot,otting2020hot,mews2021continuous}). We thus anticipate an uptake of this type of modern statistical modelling tools in future research in particular into the dynamics of live-betting markets, but also in other sports settings.

\section*{Acknowledgements}
Marius Ötting received support from the Deutsche Forschungsgemeinschaft (Grant 431536450), which is gratefully acknowledged.

\nolinenumbers  

\bibnote{luca}{[dataset]}

\bibliographystyle{apalike}
\bibliography{references}

\renewcommand{\thesection}{A}
\newpage

\section*{Appendix}

\subsection{Computation of the VAEP values}

The VAEP values are obtained using predictive machine learning as proposed in \citet{vaep}, fitting gradient-boosted trees to the data on in-game actions across five different seasons as well as one World Cup and one European Championship (data provided by \citealp{luca}). We obtain positive but small values for successful passes and dribblings in midfield, whereas corresponding actions yield larger VAEP values when occurring closer to the goal.

\subsection{Calculating and optimising the SSM likelihood}

The likelihood of our non-linear and non-Gaussian SSM involves multiple integrals, which we evaluate numerically by finely discretising the state process as first suggested by \citet{Kitagawa}. This discretisation corresponds to a reformulation of the continuous-state SSM as a discrete-state hidden Markov model (HMM) with a large state space \citep{zucchini2009hidden}.
Using the Markov property assumed for the state process and the conditional independence of the observations, given the states, the likelihood can be written as

\begin{equation}
\label{disc1}
\mathcal{L}(\theta) = f(y_1, \ldots, y_T) = \int \ldots \int f(g_1)f(y_1|g_1) \prod_{t = 2}^{T} f(g_t|g_{t-1}f(y_t|g_t)\ dg_T \ldots g_1,
\end{equation}

where the vector $\theta$ contains all model parameters. While the integrand has a simple form, the multiple integration makes this expression analytically intractable. Our approach to (approximately) evaluating the right side of Eq. (\ref{disc1}) is based on very finely discretising the state space, thus replacing the integrals by approximating sums. More specifically, for the state process $g_t$ we consider a possible range $[c_0, c_m],\ c_0, c_m \in \mathbb{R}$, chosen sufficiently large as to cover virtually all possible values the process may take. We split this range into $m$ equidistant intervals $C_i = (c_{i-1}, c_i)$, $i = 1, \ldots, m$, of length $h = (c_m - c_0)/m$, and denote the midpoint of the $i$--th interval by $c_i^{\star}$. Using quadrature with a simple midpoint rule over each of the intervals $C_1,\ldots,C_m$, expression (\ref{disc1}) is approximated by
\begin{equation}
\label{finapp}
\mathcal{L}_{\text{approx}} = h^T \sum\limits_{i_1 = 1}^m ... \sum\limits_{i_T = 1}^m f(c_{i_{1}}^{\star})f(y_1|g_1 = c_{i_{1}}^{\star}) \prod\limits_{t = 2}^{T} f(c_{i_{t}}^{\star}|g_{t-1} = c_{i_{t-1}}^{\star})f(y_t|g_t = c_{i_{t}}^{\star}).
\end{equation}
While the calculation of this expression is not computationally feasible given the large number of summands ($m^T$), this
reformulation allows us to apply recursive techniques from the HMM toolbox, in particular for calculating the approximate likelihood (for a single match).
Specifically, the approximated likelihood in (\ref{finapp}) is \textit{precisely} the likelihood of a particular $m$-state HMM. The initial state distribution of this HMM is given by the $m$-dimensional vector $\boldsymbol{\delta} = (\delta_1, \ldots, \delta_m)$ with $\delta_i = hf(c_i^{\star}).$ 
Similarly, the $m \times m$ transition probability matrix (t.p.m.) $\Gamma$ of the HMM has entries $\gamma_{ij} = h f(c_j^{\star}|c_i^{\star})$ --- the approximate probability of the state process moving from interval $C_i$ to interval $C_j$. In our model, $\Gamma$ additionally depends on covariates, such that we use the notation $\Gamma^{(t)}$ to make it explicit that the t.p.m. varies over time.
Finally, the $m$-state HMM involves the state-dependent densities $f(y_t| c_i^{\star})$, $i=1,\ldots,m$, representing the approximate density of the observation $y_t$, given that the state process $g_t$ is in the subinterval $C_i$ at time $t$. Having recognised (\ref{finapp}) as the likelihood of this particular HMM, we can use the corresponding powerful tools for inference, first and foremost the forward algorithm for efficiently evaluating (\ref{finapp}), yielding the matrix product expression
\begin{equation}
\label{unpen}
\mathcal{L}_{\text{approx}} = \boldsymbol{\delta} \mathbf{P}(y_1)\boldsymbol{\Gamma}^{(1)} \mathbf{P}(y_2)\boldsymbol{\Gamma}^{(2)} \mathbf{P}(y_3) \dots \boldsymbol{\Gamma}^{(T-2)} \mathbf{P}(y_{T-1})\boldsymbol{\Gamma}^{(T-1)} \mathbf{P}(y_T) \mathbf{1},
\end{equation}
with $\mathbf{1} = (1, \dots, 1) \in \mathbb{R}^m$. This discretisation trick allows us to calculate an arbitrarily accurate approximation of the SSM likelihood at computational cost $\mathcal{O}(m^2T)$ only (see also, for example, \citealp{bartolucci2001maximum,langrock2011some,zucchini2009hidden,mews2022}).

To evaluate the likelihood for the complete data set, independence of stakes placed across matches is assumed, such that the joint likelihood for all 306 matches is simply the product of the likelihoods of the individual matches. Parameter estimation is then carried out numerically by optimising the likelihood --- or, in case of the varying-coefficient model, the penalised likelihood --- using a Newton-Raphson-type scheme. 

\end{spacing}

\end{document}